\documentclass[10pt]{article}

\def \D {\hbox{d}}
\def      \Metric{\sigma}
\def \diag{\mathop{\rm diag}\nolimits}
\begin{document}

\title{
\textbf
 {A closed-form solution in a dynamical system related to Bianchi IX}
}

\author
{
R.~Conte\footnote{
Preprint S2007/085. To appear, Physics Letters A.
}
\\
Service de physique de l'\'etat condens\'e
(URA 2464)
\\~~CEA--Saclay, F--91191 Gif-sur-Yvette Cedex, France
\\[10pt]
\\
E-mail: Robert.Conte@cea.fr
}

\maketitle

\hfill

{\vglue -10.0 truemm}
{\vskip -10.0 truemm}

\begin{abstract}
The Bianchi IX cosmological model in vacuum
can be represented by several six-dimensional dynamical systems.
In one of them we present a new closed form solution
expressed by a third Painlev\'e function.
\end{abstract}


\noindent \textit{PACS 1995}~:
 02.30.-f,  
 05.45.+b,  
 47.27.-i,  
 98.80.Hw   

\baselineskip=12truept


\bigskip

The Bianchi IX cosmological model
in vacuum can be defined by the metric
\cite{LandauLifshitzTheorieChamps}
\begin{eqnarray}
& &
\D s^2= \Metric^2 \D t^2 - \gamma_{\alpha \beta} \D x^\alpha \D x^\beta,\
\\
& &
\gamma_{\alpha \beta}=\eta_{ab} e_\alpha^a e_\beta^b,\
\eta=\diag(A,B,C),
\end{eqnarray}
in which $e_\alpha^a$ are the components of the three frame vectors,
and
$\Metric^2=\pm 1$ according as the metric is
Minkovskian or Euclidean.
Introducing the logarithmic time $\tau$
by the hodograph transformation
\begin{eqnarray}
& &
\D \tau = \frac{\D t}{\sqrt{A B C}},
\label{eqtwotimes}
\end{eqnarray}
this gives rise to
the six-dimensional system of three second order ODEs
\begin{equation}
\Metric^2 (\log A)'' = A^2 - (B-C)^2
\hbox{ and cyclically},\
'=\D / \D \tau,
\label{eqBianchi1}
\end{equation}
or equivalently
\begin{equation}
\Metric^2 (\log \omega_1)''
= \omega_2^2 + \omega_3^2 - \omega_2^2 \omega_3^2 / \omega_1^2
\hbox{ and cyclically},\
\end{equation}
under the change of variables
\begin{eqnarray}
& &
A= \omega_2 \omega_3 / \omega_1,\
\omega_1^2=B C
\hbox{ and cyclically}.
\end{eqnarray}
\medskip

If one introduces the six variables
\begin{eqnarray}
& &
y_1=\frac{A}{\Metric},\
z_1=\frac{\D}{2 \D \tau} \log (B C),\
\hbox{ and cyclically},
\end{eqnarray}
the dynamical system (\ref{eqBianchi1})
can be alternatively represented by \cite{CGR1994}
\begin{eqnarray}
& &
\frac{\D y_j}{\D \tau}=-y_j (z_j-z_k-z_l),\
\frac{\D z_j}{\D \tau}=-y_j (y_j-y_k-y_l),
\label{eqBIXyjzj}
\end{eqnarray}
in which $(j,k,l)$ is any permutation of $(1,2,3)$.
This system admits the first integral
\begin{eqnarray}
& &
{\hskip -10.0 truemm}
\Metric^{-2} K_1=
   y_1^2+y_2^2+y_3^2- 2 y_2 y_3- 2 y_3 y_1- 2 y_1 y_2
 -(z_1^2+z_2^2+z_3^2- 2 z_2 z_3- 2 z_3 z_1- 2 z_1 z_2).
\label{eqBIXFirst}
\end{eqnarray}
\medskip

All the single valued solutions of (\ref{eqBianchi1})
are known in closed form \cite{BGPP,Taub},
except a four-parameter solution \cite{LMC1994} which would extrapolate
the three-parameter elliptic solution \cite{BGPP}
\begin{equation}
\omega_j = \Metric \sqrt{\wp(\tau-\tau_0,g_2,g_3)-e_j},\ j=1,2,3,\ K_1=0,
\label{eqEulersol}
\end{equation}
in which $\wp,g_2,g_3,e_j$ is the classical notation of Weierstrass,
\begin{eqnarray}
& &
{\wp'}^2=4 \wp^3 - g_2 \wp - g_3 = 4 (\wp-e_1)(\wp-e_2)(\wp-e_3),\
(g_2,g_3,e_j) \hbox{ complex},
\label{eqWeierstrass}
\end{eqnarray}
and $g_2,g_3,\tau_0$ are arbitrary.
The solution (\ref{eqEulersol}) also represents the motion of a rigid body
around its center of mass (Euler, 1750),
\begin{equation}
\Metric \omega_1' = \omega_2 \omega_3,
\hbox{ and cyclically}.
\label{eqEuler}
\end{equation}

Any hint to find the above mentioned missing four-parameter solution
would be welcome,
and some indications can be found in Ref.~\cite{LMC1994}.
In the present Letter,
we present such a hint,
as a five-parameter solution of (\ref{eqBIXyjzj}).
Despite its lack of physical meaning,
it could share some analytic structure with the unknown solution
and therefore provide a useful insight.
\medskip

When one coordinate $y_i$ vanishes, say $y_1=0$,
the correspondence (\ref{eqtwotimes})
between the physical time $t$ and the logarithmic time $\tau$
breaks down,
but the system (\ref{eqBIXyjzj}),
whose investigation was then started in Ref.~\cite{LlibreValls2005},
can be integrated in closed form.

Taking account of the two additional first integrals \cite{LlibreValls2005},
\begin{eqnarray}
& &
c=-z_1,\
K_2=y_2 y_3 e^{-2 c \tau},
\end{eqnarray}
the system reduces to
\begin{eqnarray}
& &
\left\lbrace
\begin{array}{ll}
\displaystyle{
(z_2-z_3)'=-(y_2+y_3) (y_2-y_3),
}\\ \displaystyle{
(y_2+y_3)'=-(y_2-y_3) (z_2-z_3) - c (y_2+y_3),
}\\ \displaystyle{
(y_2-y_3)'=-(z_2-z_3) (y_2+y_3) - c (y_2-y_3),
}\\ \displaystyle{
(z_2+z_3)'=-(y_2-y_3)^2.
}
\end{array}
\right.
\end{eqnarray}

For $c=0$,
the system for $z_2-z_3, y_2+y_3, y_2-y_3$ is another Euler top,
whose general solution is
\begin{eqnarray}
& &
\left\lbrace
\begin{array}{ll}
\displaystyle{
y_1=0,\
z_1=0,\
}\\ \displaystyle{
z_2-z_3=\sqrt{\wp(\tau-\tau_1,g_2,g_3)-e_1},\
}\\ \displaystyle{
y_2+y_3=\sqrt{\wp(\tau-\tau_1,g_2,g_3)-e_2},\
}\\ \displaystyle{
y_2-y_3=\sqrt{\wp(\tau-\tau_1,g_2,g_3)-e_3},\
}\\ \displaystyle{
z_2+z_3=    \zeta(\tau-\tau_1,g_2,g_3) + e_3 (\tau-\tau_1) + 2 z_0,\
}
\end{array}
\right.
\end{eqnarray}
in which $g_2,g_3,\tau_1,z_0$ are the four arbitrary constants.

For $c\not=0$,
the elimination of $(y_3,z_2,z_3)$ between the two first integrals
and the original system yields the general solution
\begin{eqnarray}
& &
\left\lbrace
\begin{array}{ll}
\displaystyle{
y_1=0,\
z_1=-c \not=0,\
}
\\
\displaystyle{
-y_j'' + \frac{{y_j'}^2}{y_j}
+ y_j^3 - K_2^2 e^{-4 c \tau} y_j^{-1}=0,\ j=2 \hbox{ or } 3,
}
\\
\displaystyle{
y_2 y_3 =K_2 e^{2 c \tau},
}
\\
\displaystyle{
z_2-z_3=\frac{y_3'}{2 y_3}-\frac{y_2'}{2 y_2},
}
\\
\displaystyle{
z_2+z_3=\frac{(y_2-y_3)^2 - (z_2-z_3)^2 + \sigma^{-2}K_1 - c^2}{2 c},
}
\end{array}
\right.
\label{eqPIIIsol}
\end{eqnarray}
and the second order ordinary differential equation for $y_2$
(or for $y_3$ as well)
is a third Painlev\'e equation \cite{PaiBSMF},
with the correspondence
\begin{eqnarray}
& &
\frac{\D^2 w}{\D \xi^2}=\frac{1}{w} \left(\frac{\D w}{\D \xi}\right)^2
 - \frac{\D w}{\xi \D \xi}
+ \frac{\alpha w^2 + \gamma w^3}{4 \xi^2}
+\frac{\beta}{4 \xi} + \frac{\delta}{4 w},
\\
& &
w=y_2 \hbox{ or } y_3,\
\xi=e^{-2 c \tau},\
\alpha=0,\
\beta=0,\
\gamma=1,\
\delta=-K_2^2 c^{-4}.
\end{eqnarray}
In the generic case $c K_2 \not=0$,
this solution is a meromorphic function of $\tau$,
with a transcendental dependence on the two constants of integration
other than $(c,K_1,K_2)$.


What is remarkable is that the unknown four-parameter solution
of (\ref{eqBianchi1})
and the Painlev\'e III solution (\ref{eqPIIIsol}) of (\ref{eqBIXyjzj})
are both extrapolations of an Euler top.
This suggests looking for another possible three-dimensional
Euler top in the six-dimensional physical system (\ref{eqBianchi1}).
Such a three-dimensional subsystem would necessarily correspond
to a non self-dual curvature \cite{GibbonsPope1979}.



\vfill \eject

\begin{thebibliography}{99}

\bibitem{BGPP} V.~A.~Belinskii, G.~W.~Gibbons, D.~N.~Page, and C.~N.~Pope,
Asymptotically Euclidean Bianchi IX metrics in quantum gravity,
Phys.~Lett.~A {\bf 76} (1978) 433--435.

\bibitem{CGR1994} G.~Contopoulos, B.~Grammaticos, and A.~Ramani,
The mixmaster universe model, revisited,
J.~Phys.~A {\bf 27} (1994) 5357--5361.

\bibitem{GibbonsPope1979} G.~W.~Gibbons and C.~N.~Pope,
The positive action conjecture and asymptotically Euclidean metrics in
quantum gravity,
Commun.~Math.~Phys.~{\bf 66} (1979) 267--290.

\bibitem{LandauLifshitzTheorieChamps} L.~D.~Landau et E.~M.~Lifshitz,
{\it Th\'eorie classique des champs},
chapitre ``Probl\`emes cosmologiques''
(\'Editions Mir, Moscou, 3i\`eme \'edition et suiv., 1971).

\bibitem{LMC1994} A.~Latifi, M.~Musette, and R.~Conte,
The Bianchi IX (mixmaster) cosmological model is not integrable,
Phys.~Letters A {\bf 194} (1994) 83--92; {\bf 197} (1995) 459--460.
http://arXiv.org/abs/chao-dyn/9409002.

\bibitem{LlibreValls2005} J.~Llibre and C.~Valls,
Integrability of the Bianchi IX system,
J.~Math.~Phys.~{\bf 46} (2005) 072901, 13 pp.

\bibitem{PaiBSMF} P.~Painlev\'e,
M\'emoire sur les \'equations diff\'erentielles dont l'int\'egrale
g\'e\-n\'e\-ra\-le est uniforme,
Bull.~Soc.~Math.~France {\bf 28} (1900) 201--261.

\bibitem{Taub} A.~H.~Taub,
Empty space-times admitting a three-parameter group of motions,
Annals of Math.~{\bf 53} (1951) 472--490.

\end{thebibliography}
\end{document}